\documentclass[aps,prb,twocolumn,superscriptaddress,showpacs,preprintnumbers,amsmath,amssymb,floatfix]{revtex4}
\usepackage{graphicx}
\usepackage{dcolumn}
\usepackage{bm}
\usepackage{color}
\usepackage[latin1]{inputenc}
\graphicspath{{./figs/}}

\begin{document}

\title{Signatures of sub-band excitons in few-layer black phosphorus}

\author{A. Chaves} \email{andrey@fisica.ufc.br}
\affiliation{Universidade Federal do Cear\'a, Departamento de
F\'{\i}sica Caixa Postal 6030, 60455-760 Fortaleza, Cear\'a, Brazil}
\affiliation{Department of Physics, University of Antwerp, Groenenborgerlaan 171, B2020 Antwerp, Belgium}
\author{G. O. Sousa}
\affiliation{Universidade Federal do Cear\'a, Departamento de
F\'{\i}sica Caixa Postal 6030, 60455-760 Fortaleza, Cear\'a, Brazil}
\author{K. Khaliji}
\affiliation{Electrical and Computer Engineering Department, University of Minnesota, 200 Union Street SE, 4-174 Keller Hall, Minneapolis, MN 55455-0170}
\author{D. R. da Costa}
\affiliation{Universidade Federal do Cear\'a, Departamento de
F\'{\i}sica Caixa Postal 6030, 60455-760 Fortaleza, Cear\'a, Brazil}
\author{G. A. Farias}
\affiliation{Universidade Federal do Cear\'a, Departamento de
F\'{\i}sica Caixa Postal 6030, 60455-760 Fortaleza, Cear\'a, Brazil}
\author{Tony Low}
\affiliation{Electrical and Computer Engineering Department, University of Minnesota, 200 Union Street SE, 4-174 Keller Hall, Minneapolis, MN 55455-0170}

\begin{abstract}
Recent experimental measurements of light absorption in few-layer black phosphorus (BP) reveal a series of high and sharp peaks, interspersed by pairs of lower and broader features. Here, we propose a theoretical model for these excitonic states in few-layer black phosphorus (BP) within a continuum approach for the in-plane degrees of freedom and a tight-binding approximation that accounts for inter-layer couplings. This yields excitonic transitions between different combinations of the sub-bands created by the coupled BP layers, which leads to a series of high and low oscillator strength excitonic states, consistent with the experimentally observed bright and dark exciton peaks, respectively. The main characteristics of such sub-band exciton states, as well as the possibility to control their energies and oscillator strengths via applied electric and magnetic fields, are discussed, towards a full understanding of the excitonic spectrum of few-layer BP and its tunability.

\end{abstract}
\pacs{78.66.Db 71.70.Ej 71.35.-y}

\maketitle

\section{Introduction}

Black phosphorus (BP), a stable allotrope of phosphorus composed by van der Waals stack of puckered layers, is a direct gap semiconductor. \citep{Bridgman,Keyes,Andres,Morita,Avouris,chaves2020bandgap} Over the past few years, few-layer BP has been intensively studied, especially due to the interesting properties resulting from its anisotropic band structure.\cite{PBP,LiuHan,BPfet,CastellanosGomez} The optical bandgap of BP is layer tunable, and spans over the mid-infrared to near-infrared\cite{Andres,LowT}, making it a good candidate for optoelectronics applications\cite{RuPen,YoungBlood,WW}.  

Typical bulk semiconductors like Si, Ge, and GaAs have large dielectric screening environment. As a consequence, exciton binding energy in bulk is typically negligible in comparison with thermal fluctuation, unless the measurement is made at low temperatures.\cite{JaiSingh} Conversely, the reduced dimensionality of two-dimensional (2D) materials yield strong Coulomb interaction between electrons and holes, due to the reduced dielectric screening.\cite{SLatini} In 2D materials, binding energies on the order of hundreds of meV are possible, thus allowing for measurement of exciton states at room temperature,\cite{Alexey2014, Vinod2019, GuoweiAndrey, AndreyReview} through photoluminescence (PL) and absorption spectroscopy.  

\begin{figure}[!t]
\centerline{\includegraphics[width = \linewidth]{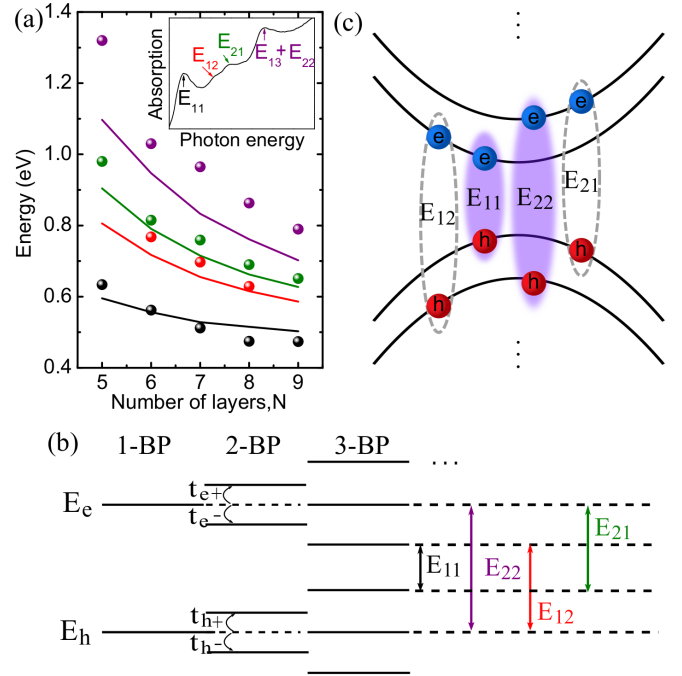}}
\caption{(color online) (a) Energy of the four low-lying experimentally observed absorption peaks \cite{AndreyTony} (symbols) as a function of the number of BP layers. An illustration of these peaks, which are theoretically assigned to transitions between conduction and valence sub-bands, is shown in the inset. Results from our model for the four low-lying sub-band excitons are shown as solid lines for comparison. (b) Diagram showing the evolution of electron and hole energy band edges as the number of stacked layers increases. The separation between sub-bands is proportional to the hopping energies $t_{e,h}$ between the coupled BP layers. (c) Sketch of the possible bright ($E_{11}$ and $E_{22}$, shaded) and dark ($E_{12}$, and $E_{21}$) sub-band excitons involved in the absorption peaks shown in the inset in (a).} \label{fig:modelsystem}
\end{figure}

Recent experiments \cite{Li2016, AndreyTony} investigated the layer dependence of optical absorption peaks in $N$-layer BP ($N$-BP). A representative absorption spectrum observed in Ref. [\onlinecite{AndreyTony}] is shown in the inset of Fig. \ref{fig:modelsystem}(a). It exhibits peaks with high intensities (labelled here as $E_{11}$ and $E_{13}+E_{22}$), accompanied with a pair of intermediate energy exciton states ($E_{12}$ and $E_{21}$), whose intensities are significantly lower. Hence, the former and latter are loosely called the bright and dark excitons respectively. More precisely, the "brightness" of these dark excitons are experimentally found to increase with the application of an electrostatic bias across the BP layers. As illustrated in Fig. \ref{fig:modelsystem}(b), the conduction and valence bands in $N$-BP are splitted into a discrete set of $N$ sub-bands. Interlayer coupling in $N$-BP is such that the conduction and valence band edges are located at the $\Gamma$-point for any number of layers, with minimal corrections even to the band curvatures and quasi-particle effective masses as $N$ increases. Hence, this material is ideal for the observation of sub-band excitons. In this paper, we present a theoretical model that properly describes exciton binding energies and their corresponding oscillator strengths in $N$-layer BP, as well as their dependence on external parameters such as different dielectric environment and applied electric and magnetic fields.

\section{Theoretical model}

We consider a simple theoretical model that employs effective mass approximation (EMA) for the in-plane coordinates, along with a tight-binding (TB) model to account for the coupling between stacked BP layers. Our theoretical model successfully capture the basic features of the dark and bright exciton states, and the experimentally observed brightening of dark excitons in the presence of a vertical bias \cite{AndreyTony}. The dependence of the electron-hole overlap in excitonic transitions, as an indirect measure of their oscillator strength, on the number of layers and applied field intensity is addressed as well. Different sub-band states are also demonstrated to exhibit different electric dipoles and excitonic radii. 

In general, for an arbitrary three dimensional (3D) semiconductor, the electron-hole Hamiltonian within the EMA reads
\begin{equation}
H = \sum_{i=e,h} \left[-\frac{\hbar^2}{2m_i^{\parallel}}\nabla_{\parallel,i}^2 - \frac{\hbar^2}{2m_i^{\perp}}\nabla_{\perp,i}^2 + V_i(z_i)\right] + V(\vec{r}_e-\vec{r}_h)
\end{equation}
where the effective mass is assumed to be different in the in-plane ($m_i^{\parallel}$) and out-of-plane ($m_i^{\perp}$) directions, and the index $i = e(h)$ stands for electron (hole). Quasi-particles are assumed to be under a potential $V_i(z_i)$ (e.g., due to a perpendicularly applied electric field, or even a possible inter-layer band offset) and to interact via the potential $V(\vec{r}_e-\vec{r}_h)$. 

The in-plane coordinates are re-written in terms of relative ($\vec{\rho} = \vec{\rho}_e - \vec{\rho}_h$) and center-of-mass ($\vec{R} = (m_e^{\parallel}\vec{\rho}_e+m_h^{\parallel}\vec{\rho}_h)/(m_e^{\parallel} + m_h^{\parallel})$) coordinates. The kinetic energy terms in the Hamiltonian related to the exciton center-of-mass can be ignored, as they are good quantum numbers in a translational invariant system. In the case of $N$-layer BP, the out-of-plane direction does not have translation symmetry, hence it requires real space representation in terms of discrete electron and hole $z_i$ coordinates. The Schr\"odinger equation for Wannier excitons in $N$-layer BP under a perpendicular electric field $F$ now reads 
\begin{eqnarray}
H_{exc}\Psi_{ij}(\vec\rho) = \left[-\frac{\hbar^2}{2\mu^{x}_{ij}}\frac{\partial^2}{\partial x^2}-\frac{\hbar^2}{2\mu^{y}_{ij}}\frac{\partial^2}{\partial y^2} + V_{ij}(\rho)\right]\Psi_{ij}(\vec\rho) \nonumber \\ - t_e\left[\Psi_{i+1~j}(\vec\rho)+\Psi_{i-1~j}(\vec\rho)\right]  \nonumber \\
- t_h\left[\Psi_{i~j+1}(\vec\rho)+\Psi_{i~j-1}(\vec\rho)\right] + eFd(i-j)\Psi_{ij}(\vec\rho)
\end{eqnarray} 
where $i(j) = 1, 2, 3, ... N$ represent the layer occupied by the electron (hole), the reduced effective mass $1/\mu^{x(y)}_{ij} = 1/m_{e i}^{x(y)} + 1/m_{h j}^{x(y)}$ explicitly takes into account the anisotropy of the electron and hole effective masses in the $x$ (armchair) and $y$ (zigzag) directions, and $d$ is the inter-layer distance (assumed to be the same as the layer thickness). For the perpendicular direction, $t_{e(h)}$ is the hopping parameter coupling the BP layers. 

We assume a low free carrier density ($<$10$^{12}$ cm$^{-2}$), so that screening effects on the electric field potential\cite{PBP} across the layers can be neglected. \cite{Partoens, DArf} The interaction potential $V_{ij}(\rho)$ between an electron in the $i$-th BP layer and a hole in the $j$-th layer is obtained in a semi-analytical way by generalizing the Rytova\cite{Rytova} and Keldysh \cite{Keldysh} solution of Poisson equation for a stack of slabs representing the substrate, few-layer BP, and superstrate. This potential reduces to the Rytova-Keldysh potential \cite{Rytova, Keldysh} in the zero thickness limit, whereas in the bulk limit, it assumes the Coulomb form. Such a potential has been recently developed for the general case of van der Waals heterostructures in Ref. [\onlinecite{Lucas2017}]. In the Appendix, we adapt it to the specific case of multi-layer BP, and the resulting potential will be used in the calculations that follow. As expected, in the bulk limit, this potential assumes the Coulomb form $V_{ij}(\rho) = V_{C}(\sqrt{\rho^2 + (z_e - z_h)^2})$, where $z_{e(h)} = i(j)d$, as demonstrated in Ref.[\onlinecite{Lucas2017}]. The Hamiltonian $H_{exc}$ is general and can accommodate heterostructures of different 2D semiconducting materials as well: in this case, one must assume layer dependent effective masses, include band offsets in the quasi-particle potential and adjust the inter-layer hopping terms $t_i$ accordingly.   

From now onwards, unless otherwise explicitly stated, we will assume a sample encapsulated by bulk hexagonal BN (hBN), since many samples in the literature are made as such. This is mostly due to the fact that few-layer BP, when exposed to air, reacts with the environment and changes its properties with time.\cite{Island2015, Artel2017} Encapsulation with hBN will affect the dielectric environment and, consequently, the strength of electron-hole interactions. The inclusion of such dielectric environment in the effective electron-hole potential is discussed in details in the Appendix.

In the basis $(\Psi_{11} ~ \Psi_{12} ~ ... ~\Psi_{1N} ~ \Psi_{21} ~ \Psi_{22} ~... \Psi_{NN})$, the exciton Hamiltonian is rewritten as a block matrix
\begin{equation}\label{eq.excHamfull}
H_{exc} = \left( \begin{tabular}{ccccc}
A$_1$ & B & 0 & 0 & ... \\
B & A$_2$ & B & 0 & ... \\
0 & B & A$_3$ & B & ... \\ 
0 & 0 &$\ddots$ & $\ddots$ & $\ddots$ \\ 
0 & 0 & ... & B & A$_N$ \\ 
\end{tabular} \right)
\end{equation}
where the 0's stand for $N \times N$ blocks of null matrices, $B$ are $N \times N$ diagonal matrices whose diagonals are $-t_e$, and $A_i$ are $N \times N$ tridiagonal matrices
\begin{widetext}
\begin{equation}\label{eq.Aterms}
A_i = \left( \begin{tabular}{ccccc}
$H_{i1}+ eFd(i-1)$ & $-t_h$ & 0 & 0 &  ... \\
$-t_h$ & $H_{i2} + eFd(i-2)$ & $-t_h$ & 0 & ...\\
0 & $-t_h$ & $H_{i3} + eFd(i-3)$ & $-t_h$ & ...\\
0 & 0 & $\ddots$ & $\ddots$ & $\ddots$ \\ 
0 & 0 & ... & $-t_h$ & $H_{iN}+ eFd(i-N)$
 \end{tabular} \right),
\end{equation}
\end{widetext}
with 
\begin{equation}
H_{ij} =  -\frac{\hbar^2}{2\mu^{x}_{ij}}\frac{\partial^2}{\partial x^2}-\frac{\hbar^2}{2\mu^{y}_{ij}}\frac{\partial^2}{\partial y^2} + V_{ij}(\rho). 
\end{equation}

Our model does not take into account the possible modification of the band curvatures (and, consequently, effective masses) along in-plane directions as the number of layers increase. The effect of stacking BP layers in our model modifies the electron and hole energy levels, but not their effective masses. Nevertheless, effective masses in BP are shown not to change significantly for $N \geq5$.\cite{Duarte2017} Therefore, in this work we investigate only systems with 5 layers or more, so that we can describe it with the same $t_{e(h)}$ and in-plane effective masses.

\begin{figure}[!b]
\centerline{\includegraphics[width =\linewidth]{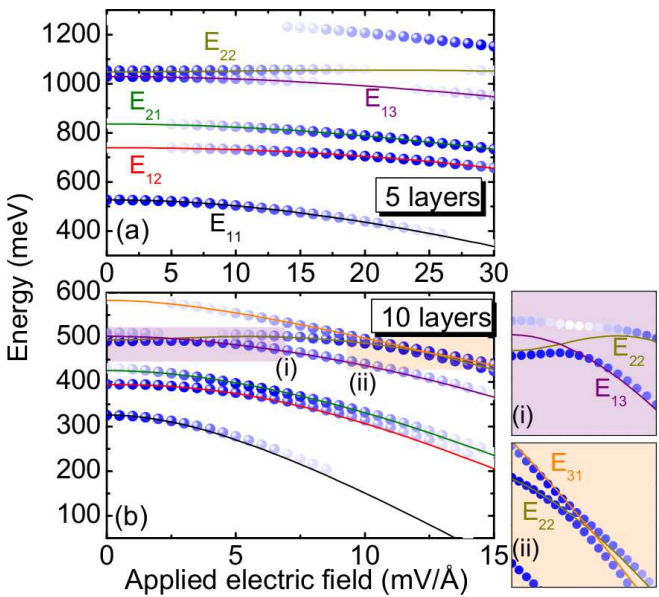}}
\caption{Lowest energy exciton levels in BP with $N$ =  5 (a) and 10 layers (b) as a function of the applied electric field. Darker symbols represent stronger electron-hole overlap. Solid lines and labels represent the sub-band transitions, in the absence of electron-hole interactions, associated with each exciton energy state, hard shifted down in energy by an amount equal to the ground state binding energy of the interacting system (symbols), for comparison. Results in the shaded regions labelled (i) and (ii) in (b) are magnified in the side panels, where lines representing non-interacting electron-hole states exhibit crossings, in contrast to the symbols, for interacting electrons and holes, where anti-crossings are observed.} \label{fig:EnergyvsField}
\end{figure}

\begin{figure*}
\centerline{\includegraphics[width =0.7\textwidth]{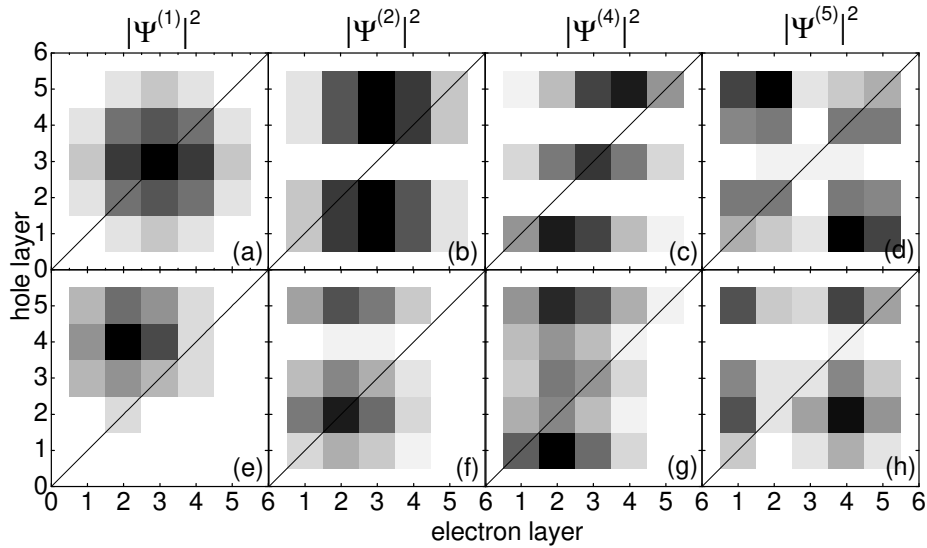}}
\caption{Grayscale map of the probability densities of the $n = 1, 2, 4$ and 5 excitonic eigenstates $\Psi^{(n)}$ of 5-BP (a-d) in the absence of applied electric fields, and (e-h) under a $F = 20$ mV/\AA\, field. Darker color represents higher probability density. The predominant sub-band exciton character of each state is shown in Fig. \ref{fig:EnergyvsField}(a). The $n$ = 3 state is qualitatively similar to the $n = 2$ and is thus omitted here for conciseness. The solid line is a guide to the eye, representing a situation where electron and hole are in the same layer, leading to maximum oscillator strength.} \label{fig:WF5}
\end{figure*}

Discretization of the in-plane coordinates into a grid with $N_x \times N_y$ points in a finite difference scheme would allow one to perform full diagonalization of the Hamiltonian of Eq. (\ref{eq.excHamfull}), which then yields exact eigenvalues of the 3D problem proposed here. This may be convenient for a small number of layers with isotropic properties, where one may convert the $N_x \times N_y$ into a smaller $N_r$ points grid along the radial coordinate. However, as $N$ increases and anisotropic effective masses are assumed, the $N^4$ matrix Hamiltonian would become a $N_x \times N_y\times N^4$ matrix which, although sparse, may be computationally expensive to be numerically diagonalizable. In order to overcome this problem, as an approximation, we assume the exciton Wannier equation for in-plane coordinates is separable from the out-of-plane, i.e., $H_{ij}\Psi_{ij} = E^{2D}_{ij}\Psi_{ij}$. All the $H_{ij}$ terms in Eq. (\ref{eq.Aterms}) are then replaced just by the $E^{2D}_{ij}$ energy of the 1s state obtained from numerical solution of the in-plane Schr\"odinger equation for the Hamiltonian $H_{ij}$.\cite{AndreyRC} Most of the computational costs are attributed to this part of the procedure, where the 2D Schr\"odinger equation is systematically solved for every combination of $i$ and $j$. Only the 1s exciton state is taken into account in the calculations, as this is the most prominent feature in PL and reflectance experiments.\cite{AndreyTony} Since the in-plane and out-of-plane coordinates are decoupled, the exciton Rydberg series would just be superimposed over the exciton eigenstates of $H_{exc}$ as calculated by our method. The proposed approach has similarly been adopted in the past decade for solving quantum well problems in semiconductor heterostructures. \cite{QW1, QW2} 

The inter-layer hopping parameters $t_{e(h)}$ must be so that, as one stacks $N$ BP layers, the quasi-particle kinetic energies obtained from $H_{exc}$ in the absence of electron-hole interactions match conduction and valence band edges as obtained from TB model or density functional theory (DFT) calculations. Moreover, as the number of layers $N$ increases further, sub-band electron (hole) energy states get closer, until they merge into a continuous conduction (valence) band, which represents the band structure of bulk BP along the out-of-plane direction. In this bulk limit, the sub-band states are no longer distinguishable as separate peaks in absorption spectra, as experimentally demonstrated for $100$ nm-thick BP in Ref. [\onlinecite{AndreyTony}]. Therefore, $t_e$ and $t_h$ must also be such that the curvature of their corresponding bands in the $N \rightarrow \infty$ limit are compatible to the electron and hole effective masses along the out-of-plane direction of bulk BP.\cite{Morita} The procedure to obtain appropriate hopping parameters under these conditions, as well as the $N \rightarrow \infty$ (bulk) limit in our model, are thoroughly discussed in the Appendices.

\section{Results and discussion}
\subsection{Bright and dark sub-band excitons} 

Exciton energy states are obtained as eigenvalues of $H_{exc}\Psi^{(n)} = \varepsilon_n \Psi^{(n)}$, where  indexes $i$ and $j$ in the components $\psi^{(n)}_{ij}$ of the $n$-th state eigenfunction $\Psi^{(n)}$  stand for the layers over which electrons and holes are distributed, respectively. The lowest four eigenenergies of the exciton Hamiltonian $H_{exc}$ as a function of the number of layers are shown as solid lines in Fig. \ref{fig:modelsystem}(a), which exhibit good qualitative (and nearly quantitative) agreement with reported experimental results (symbols).\cite{AndreyTony} Each eigenstate of $H_{exc}$ represents an exciton, with energy $\varepsilon_n$, given by a linear combination of \textit{sub-band states} associated with transitions between valence and conduction sub-bands, with indices $v$ and $c$, respectively, where $v,c = 1, 2, ... N$, as illustrated by the $E_{vc}$ labels in Fig. \ref{fig:modelsystem}(c). The electron-hole (e-h) overlap is represented as the square modulus of the exciton wave function at the origin of the e-h relative coordinates system, $|\Psi(|\vec r| = 0)|^2$. This quantity is a relevant component of the oscillator strength, which dictates the intensity of absorption peaks in e.g. Elliot theory. \cite{Elliot}  In the context of the discrete Hamiltonian Eq. (\ref{eq.excHamfull}), this overlap is calculated as $|\psi^{(n)}_{11}|^2 + |\psi^{(n)}_{22}|^2+...+|\psi^{(n)}_{NN}|^2$, i.e. it is a combination of all contributions from the $i = j$ terms of the excitonic probability density for a given $n$-th eigenstate of $H_{exc}$. 

\begin{figure*}
\centerline{\includegraphics[width =0.7\textwidth]{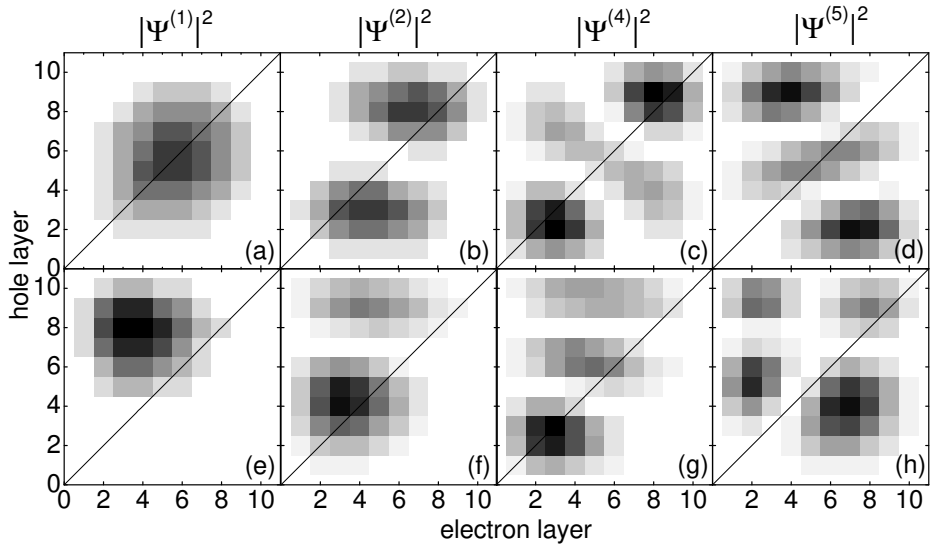}}
\caption{Same as Fig. \ref{fig:WF5}, but for 10-BP and $F$ = 5 mV/\AA\,. The predominant sub-band exciton character of each state is shown in Fig. \ref{fig:EnergyvsField}(b).} \label{fig:WF10}
\end{figure*}

Figure \ref{fig:EnergyvsField} shows the lowest exciton energy levels in BP with $N = 5$ and 10 layers as a function of the perpendicularly applied electric field. Darker symbols represent higher electron-hole overlap and, consequently, higher oscillator strength. Model parameters used in this calculation are provided in Ref. [\onlinecite{Rudenko2015}] and in the Appendix. Strong fields are expected to change the band structure of few-layer BP, \cite{Kim2015} namely, by closing its gap, therefore, we restrict our calculations to lower fields. The ground state energy is demonstrated to be reduced by up to $\approx$ 100 meV for a 30 mV/\AA\, field in 5-layer BP. Thicker samples are more sensitive to the applied field, so that lower fields are needed to tune the ground state energy in this case. For 10-layer BP, for example, $\approx$ 80 meV reduction in the ground state energy is already obtained with a field as low as 5 mV/\AA\,, and hybridization of higher energy sub-band states is observed at 2.5 mV/\AA\, and 12 mV/\AA\, fields, as we will discuss further, in the next Section. The applied field also significantly reduces the electron-hole overlaps of the ground state exciton transitions, which drop to $\approx$ 38$\%$ and $\approx$ 10$\%$ of their zero field values in the former and latter cases, respectively. Solid lines represent the energy states in the absence of electron-hole interaction, hard shifted down by an amount equal to the ground state exciton binding energy in each case, namely, $E_{11}^b = 78$ meV and 75 meV for the $N$ = 5 and $N$ = 10 cases, respectively.

\begin{figure}[!b]
\centerline{\includegraphics[width =1.0\linewidth]{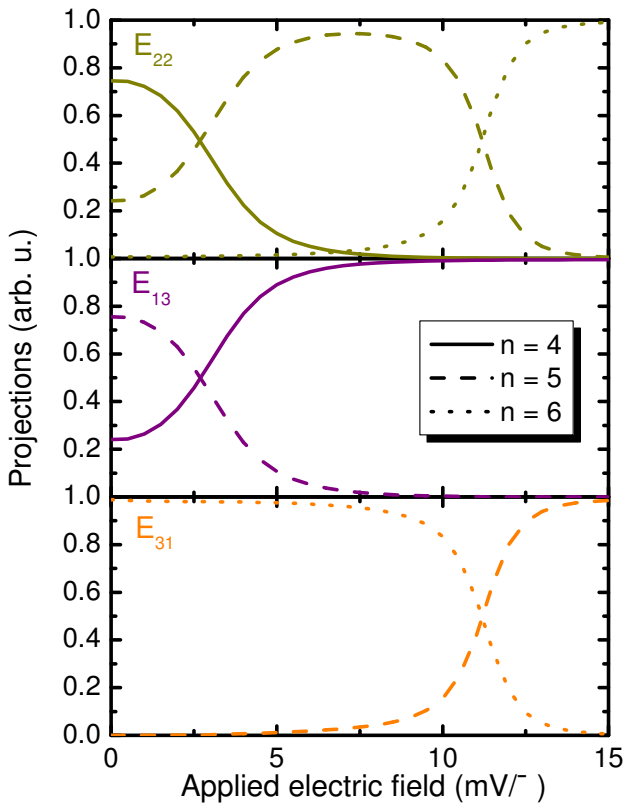}}
\caption{Projections of the $n = 4$ (solid), 5 (dashed) and 6 (dotted) exciton states $\Psi^{(n)}$ of 10-BP on the sub-band states $E_{22}$ (top), $E_{13}$ (middle) and $E_{31} $ (bottom), as a function of the applied electric field.} \label{fig:Projs}
\end{figure}

\begin{figure}[!t]
\centerline{\includegraphics[width =0.9\linewidth]{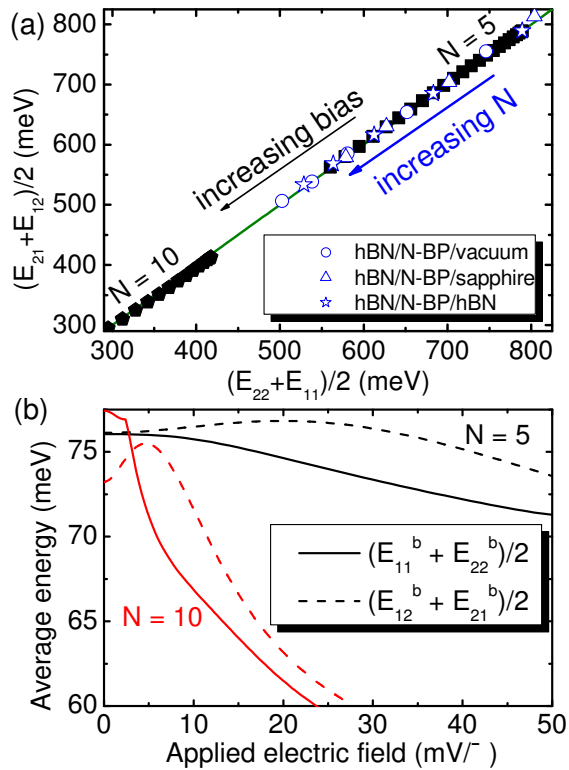}}
\caption{(color online) (a) The average between binding energies of bright (solid) and dark (dashed) sub-band exciton states as a function of the electric field for $N = 5$ (black) and 10 (red) BP layers. Difference between averages in each case is of a few meV only. (b) The average between the energies of dark exciton peaks is shown as a function of the average between the bright exciton energies (symbols). The green line is a guide to the eye, representing a situation where these two averages are equal. Different values are results for systems with different numbers of layers $N = 5, 6 ... 10$, under different dielectric environments (blue open symbols) and zero field, or under applied fields from 0 to 50 mV/\AA\, for $N = 5$ and 10 in the hBN-encapsulated case (black full symbols).} \label{fig:Averages}
\end{figure}

Notice that for $N = 5$ in Fig. \ref{fig:EnergyvsField}(a), in the absence of electric field, the ground, third and fourth excited states exhibit high oscillator strength, while the first and second excited states are found to be optically inactive. However, these states become optically active as the applied field increases. This corroborates with the experimental observations in Ref. [\onlinecite{AndreyTony}], where two weak peaks were observed with energies in between two stronger peaks in the absorption spectra of few-layer BP samples. The weak peaks in Ref. [\onlinecite{AndreyTony}] were attributed to unintentional dopings variations in the sample, since no electric field was applied in the experiment. This was confirmed by the fact that intentionally increasing the surface doping in the experiment leads to higher intensity of the weak peaks, just as one can conclude from Fig. \ref{fig:EnergyvsField}, for the analogous case of increasing the perpendicularly applied electric field.

The electric field control of the different oscillator strengths of the exciton states $\Psi^{(n)}$ can be better understood by analyzing the probability densities of these states, shown in Figs. \ref{fig:WF5} and \ref{fig:WF10}, for 5-BP and 10-BP, respectively. The diagonal solid line in each panel represents the situation where both electron and hole are in the same layer. In 5-BP, in the absence of fields (see Fig. \ref{fig:WF5}(a-d)), the exciton ground state probability density $|\Psi^{(1)}|^2$ is strongly concentrated with both electron and hole in the third (central) layer. In contrast, the probability densities of $n = 2$ and $n = 3$ (not shown) have their maxima away from the diagonal solid line. For $n = 2$, the electron wave function has a maximum at the third layer, whereas the hole wave function has a node at the same layer, thus reducing the overall oscilator strength. The probability density for $n = 3$ is qualitatively similar to that of $n = 2$, but with the maximum of the hole wave function at the third layer, where the electron wave function has a node. Since the results for $n = 2$ and $n = 3$ are similar, only the former is shown in Figs. \ref{fig:WF5} and \ref{fig:WF10}, for the sake of conciseness. For $n = 4$, the exciton proability density maximum is again at the third layer. Although the maxima for $n = 5$ are away from the diagonal line, this line still hosts a significant part of the probability density, which makes the oscillator strength of this state still relatively high in the absence of fields.

As the $F = 20$ mV/\AA\, is applied to 5-BP, see Fig. \ref{fig:WF5}(e-h), the maxima of the probability densities of $n = 1$, 4 and 5 are pushed away from the diagonal line, yielding lower oscillator strength for these states, whereas for $n = 2$ and 3, they lay exactly at the diagonal. For instance, $n = 2$ in Fig. \ref{fig:WF5}(f) shows both electron and hole concentrated at the second layer, thus increasing the oscillator strength of this state under such an applied field.

Similarly, the probability density for $n = 1$ exciton in 10-BP in the absence of field is shown in Fig. \ref{fig:WF10}(a), exhibiting a 1s-wave profile for the exciton wave function along the relative $z$-direction, with the electron and hole concentrated in the central layers. A p-wave-like doublet is observed for $n = 2$ (see Fig. \ref{fig:WF10}(b)) and 3, thus leading to lower oscillator strength for these states, whereas $n = 4$ and 5 are 2s-like states, thus, bright. In the presence of a $F = 5$ mV/\AA\, field, probability density maxima are pushed away from the diagonal for $n = 1$, 4 and 5, whereas it gets more concentrated around the third layer for $n = 2$, thus increasing the oscillator strength of this state.  

The electron-hole separation of the sub-band exciton states along the out-of-plane direction in the presence of the electric field is discussed in details in the Appendix. This separation has an upper bound determined by the $N$-BP thickness, which is $Nd = 26$ \AA\, (52 \AA\,) for $N = 5$ (10). For the exciton ground state, we observe an electron-hole separation as high as $\approx$ 17 \AA\, ($\approx$ 32.5 \AA\,) at a 90 mV/\AA\, (18 mV/\AA\,) field.    

Notice that the number of layers $N$, which determines the thickness of the system, imposes a boundary condition on the exciton wave function along the out-of-plane $z$-direction. The s-wave and p-wave profiles observed in Fig. \ref{fig:WF10}(a-d) can only be formed if the thickness of the sample is larger than the effective Bohr radius of excitons along $z$. The effective Bohr radius can be estimated as $a_{eff} \approx a_B \epsilon/\mu_z = 45.8$ \AA\,, where $a_B$ is the Bohr radius, $\epsilon = 10 \epsilon_0$ is the effective dielectric constant of BP, and $\mu_z = 0.115 m_0$ is the reduced effective mass along the $z$-direction. This condition is met for 10-BP, where the thickness is 52 \AA\,$> a_{eff}$, but not for 5-BP, where the thickness is 26 \AA\, $< a_{eff}$. Therefore, for 5-BP, the effect of boundary conditions is relevant and, consequently, s-wave and p-wave profiles are not clear in the probability density distributions in Figs. \ref{fig:WF5}(a-d).          
 
We point out that the pattern of experimentally observed\cite{AndreyTony} high and low intensity spectral features due to sub-band excitons discussed here is significantly different from the absorption features observed in other semiconductor 2D homobilayers, e.g. in bilayer molybdenum- and tungsten-based transition metal dichalcogenides (TMD). This is due to other more relevant effects in the interlayer coupling in these TMD, which induces, for instance, a shift in reciprocal space from the conduction band minimum at the K-point in the monolayer case to a satellite Q-point band edge in the bilayer case, so that the quasi-particle gap becomes indirect in the latter.\cite{molas2017, zhao2013, mak2010}

\subsection{Hybridization of sub-band excitons}

The existence of high oscillator strength excitons, whose electron-hole overlap decreases with the applied field, interspersed by a pair of low oscillator strength excitons, where the overlap rather \textit{increases} with the same field, is a clear signature of sub-band excitons. The former (latter) are associated with the $E_{11}$, $E_{22}$ and $E_{13}$ ($E_{12}$ and $E_{21}$) sub-band excitons sketched in Fig. \ref{fig:modelsystem}. This is confirmed also by a projection of the eigenstates of $H_{exc}$ over the sub-band states. In fact, in the absence of electron-hole interactions, the eigenstates of $H_{exc}$ are exactly the sub-band transitions, whereas the presence of the interacting potential mixes the sub-band exciton states in the eigenstates of $H_{exc}$, especially for higher layer numbers, where eigenstates are closer and thus more prone to energy crossing. Therefore, projection over the sub-band states reveal that the first five eigenstates of $H_{exc}$ in Fig. \ref{fig:EnergyvsField}(a), for $N = 5$, are \textit{predominantly} $E_{11}$, $E_{12}$, $E_{21}$, $E_{13}$ and $E_{22}$, respectively, from lowest to highest energy. For $N = 10$, the fourth state becomes $E_{22}$, whereas the fifth state is $E_{13}$. The latter anti-crosses with $E_{22}$ at $\approx$ 2.5 mV/\AA\, field, as shown in the side plot (i) of Fig. \ref{fig:EnergyvsField}(b), which magnifies the results in the shaded area of same label in this panel. An anti-crossing between $E_{22}$ and $E_{31}$ also occurs at $\approx$ 12 mV/\AA\, field, as shown in the side plot (ii). In the absence of e-h interaction, inter-sub-band mixing is forbidden and both anti-crossings become crossings, as shown by the solid lines in the side plots.

The hybridization of sub-band states in the $n = 4$ (solid), 5 (dashed) and 6 (dotted) excited states of the exciton in 10-BP is better illustrated in Fig. \ref{fig:Projs}, which shows the interplay between $E_{22}$, $E_{13}$ and $E_{31}$ sub-band states in the composition of these excitonic levels as the applied electric field increases. As previously discussed, the third excited excitonic state ($n = 4$) is a predominantly $E_{22}$ sub-band state (see top panel) in the absence of fields, which turns into a $E_{13}$ sub-band state for fields higher than $\approx$ 2.5 mV/\AA\, (see middle panel). For $F > 2.5$ mV/\AA\,, the $n = 5$ exciton state acquires the $E_{22}$ character up to $F \approx$ 12 mV/\AA\,, where the $n = 6$ exciton state becomes the $E_{22}$ sub-band state, while $n = 5$ acquires the $E_{13}$ character (see bottom panel).    

\subsection{Sub-band excitons Stark shift}

As illustrated in the sketch in Fig. \ref{fig:modelsystem}(b-c), the sub-bands are displayed such that one can also infer that, in the absence of e-h interaction, the average of $E_{11}$ and $E_{22}$, and that of $E_{21}$ and $E_{12}$, are equal, even in the presence of an electric field. This is clear as the field induced shift, in this case identified as a Franz-Keldysh shift of the sub-bands, would be given as a sum of the shifts in conduction ($\Delta E^{i}_{c,FK}$) and valence ($\Delta E^{j}_{v,FK}$) sub-bands ($i, j = 1, 2$) as
\begin{equation}\label{eq.FKandS}
\Delta E^{ij}_{FK} = \Delta E^{i}_{c,FK} + \Delta E^{j}_{v,FK}.
\end{equation}
The presence of e-h interactions includes an extra exciton Stark shift term, accounting for binding energies, $\Delta E^{b}_{ij}$ in Eq. (\ref{eq.FKandS}). Since, in principle, $\Delta E^{b}_{11} + \Delta E^{b}_{22} \neq \Delta E^{b}_{12} + \Delta E^{b}_{21}$, the above mentioned rule of equal averages, $(E_{11}+E_{22})/2 = (E_{12}+E_{21})/2$, might not apply.

Results for hBN encapsulated $N$-BP are shown in Fig. \ref{fig:Averages}(a) for different applied fields, from 0 (higher energy values) up to 50 mV/\AA\, (lower energy values), for $N = 5$ and 10 (full black symbols). The role of the dielectric environment in the rule of equal averages is also investigated in Fig. \ref{fig:Averages}(a): the substrate is considered to be hBN, while different symbols refer to different superstrates (open blue symbols) at zero field and for $N = 5, 6,.. 10$. The effect of the dielectric environment on the effective electron-hole interaction potential is discussed in details in the Appendix. A closer inpection into the binding energies of these exciton states in Fig. \ref{fig:Averages}(b) reveal that the average of $E_{11}$ and $E_{22}$ sub-band binding energies (solid) differs from that of $E_{12}$ and $E_{21}$ sub-band states (dashed) for $N = 5$ (black) and 10 layers (red) by a few meV, although in the former case they are accidentally similar exactly at zero field. As a consequence, even accounting for exciton Stark shifts, the rule inferred here still seems robust with respect to layer number and applied fields, as it has been experimentally probed in Ref. [\onlinecite{AndreyTony}], but deviations of the order of meV can be observed due to the exciton stark shift.  

\begin{figure}[!t]
\centerline{\includegraphics[width =0.9\linewidth]{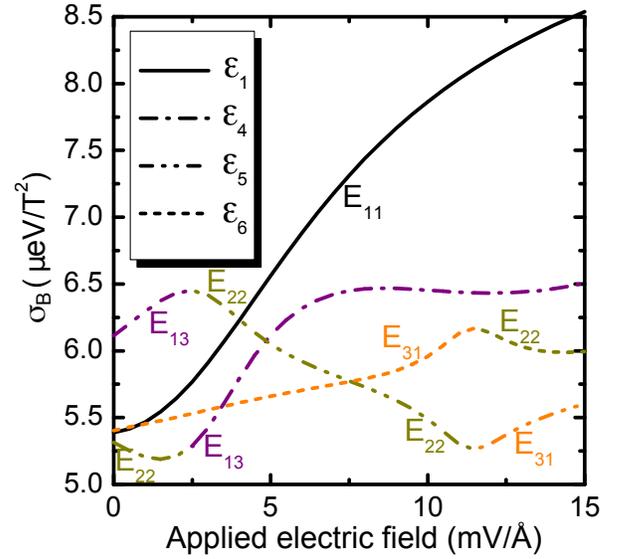}}
\caption{(color online) Diamagnetic shift factors of four low lying energy states in $10$-BP, namely, the ground state (solid) and the fourth (dashed dotted), fifth (dashed dotted-dotted) and sixth (dashed) excited states, as a function of the applied electric field intensity.  The predominance of the $E_{11}$ (black), $E_{22}$ (yellow), $E_{13}$ (purple), or $E_{31}$ (orange) sub-band states in each of these eigenstates depends on the the applied field.} \label{fig:Diamagnetic}
\end{figure}

\subsection{Diamagnetic shift of excitons}

With the sub-band exciton eigenstate assessment of each absortion peak, one can also predict the bias dependence of the average values $\langle x^2 \rangle$ and $\langle y^2 \rangle$, which are related to the diamagnetic shift $\Delta E_{dia} = \sigma_B B^2$ of these peaks under an applied magnetic field as 
\begin{equation}
\sigma_B = \frac{e^2}{8}\left( \frac{\langle x^2 \rangle}{\mu_y} + \frac{\langle y^2 \rangle}{\mu_x} \right).
\end{equation}

The diamagnetic shift factors $\sigma_B$ for 10-BP encapsulated by hBN, shown in Fig. \ref{fig:Diamagnetic}, are an order of magnitude higher than those observed for 1s excitons in monolayer TMD \cite{Stier2016, Stier2018, Zipfel2018, Goryca2019, Chen2019Luminescent}, due to the lower exciton binding energy resulting from the dielectric screening in the thicker 10-layer BP considered here. In TMD, the Rydberg series of exciton peaks in absorption \cite{Stier2016, Stier2018, Zipfel2018, Goryca2019} and PL \cite{Chen2019Luminescent} spectra exhibits diamagnetic shift factors $\sigma_B$ that consistently increase from $\approx$0.5 $\mu$eV/T$^2$, for 1s states, to $\approx$20$\mu$eV/T$^2$ for 3s states.\cite{Stier2018,Chen2019Luminescent} This is a consequence of the increasing excitonic Bohr radius and decreasing binding energies from 1s to 3s states. 

Conversely, series of sub-band exciton states in Fig. \ref{fig:EnergyvsField} is expected to exhibit diamagnetic shift factors $\sigma_B$ that are only slightly different for each sub-band exciton peak, due to their similar binding energies but different intra- and inter-layer exciton composition. By controlling the layer distribution of electron and hole wave functions, an applied electric field can be used to tune the diamagnetic shift factor. Indeed, Fig. \ref{fig:Diamagnetic} shows that an applied electric field leads to a change in the order of diamagnetic shift values of the sub-band exciton states: the ground state (black solid) increases and becomes the most susceptible state to magnetic fields. Moreover, the anti-crossings between  $E_{13}$ and $E_{22}$, and between $E_{22}$ and $E_{13}$ sub-band states, observed in Fig. \ref{fig:EnergyvsField}(b) at $F \approx 2.5$ mV/\AA\, and $\approx$ 12 mV/\AA\, fields, respectively, show up as bumps in the diamagnetic shifts of these states. Hence, the diamagnetic shifts can be an experimental probe to reveal signatures in the inter-sub-band crossing.

\section{Conclusions}

In summary, we have demonstrated that the physics of sharp excitonic peaks observed in recent PL experiments with few-layer black phosphorus is dominated by sub-band excitons. Results from our model explain the existence of a pair of intermediate dark exciton states spectrally observed in between the two lowest energy bright excitonic transitions in few-layer BP: \cite{AndreyTony, GuoweiAndrey} dark (bright) excitonic states are interpreted in terms of transitions between states in conduction and valence sub-bands with different (the same) indices. Besides, our knowledge on the sub-band exciton character of these peaks allow us to predict that the brightness of such dark states can be controlled either by an external field, or by different substrates and capping layers. Intermediate cross-sub-band states become more prominent as the applied electric field increases. Conversely, the electron-hole overlaps of the bright states are reduced by the increasing field, thus suggesting longer exciton lifetimes in biased BP stacks.\cite{yang2015, Robert2016, Massicote2016} Our study laid the groundwork for future exploration of subband excitons in multilayer vdW materials.

\acknowledgements This work was supported by the Brazilian Council for Research (CNPq), through the PRONEX/FUNCAP, Universal, and PQ programs, and by the Flemish Research Foundation (FWO).

\section*{Appendix: Electron-hole interaction potential in few-layer black phosphorus}

We show the steps towards an analytic expression for the electron-hole interaction potential in an environment consisting of $N$ stacked layers of the same material, encapsulated in an environment with arbitrary dielectric constant. The method used here is based on the electrostatic transfer matrix (ETM)\cite{Lucas2017}, and it supports the calculation of the potential considering electron and hole in different layers, that is addressed as an inter-layer exciton. Notice this situation is not supported by the Rytova-Keldysh potential\cite{Rytova,Keldysh}, where electrons and holes are supposed to be in the same layer and approximations on the layer thickness and dielectric constants are taken, which thus leads to a dielectric function that is linear in the reciprocal space variable $k$. One can verify that the Rytova-Keldysh linear form is an approximate particular case of the potential derived here.

Consider each BP layer as a slab with dielectric constant $\epsilon_2$ and thickness $L$, given by the interlayer distance between adjacent BP layers. We restrict the charge positions to the middle of each slab, with vertical coordinate $z_i$ ($i = 1, 2, ... N$) as illustrated in Fig. \ref{fig:modelsystem}. The potential $\Phi_{n}$ generated along one of the BP layers by a charge placed in a layer $n$ is found by solving the Poisson equation
\begin{equation}\label{poisson}
\epsilon_{n}^{\parallel}\nabla_{\rho,\theta}\Phi_{n} + \epsilon_{n}^{\perp}\frac{\partial^2 \Phi_{n}}{\partial z^2} = -e\delta(\vec{r}).
\end{equation}
This right hand side (RHS) of this equation holds only in the BP region, where the charge is placed. In the upper and lower h-BN layers (see Fig. \ref{fig:modelsystem}), the RHS term is zero and one has a Laplace equation.

For the sake of simplicity, let us consider $\epsilon_n^{\parallel} = \epsilon_n^{\perp} = \epsilon_n$, and take $\epsilon_n = \sqrt[3]{\epsilon_x\epsilon_y\epsilon_z}$.\cite{CastellanosGomez} The solution to Eq. (\ref{poisson}) at arbitrary $z$ for a particle in the $n$-th BP layers is given by:
\begin{widetext}
\begin{equation}\label{solution}
\Phi_{n}(\rho,z) = \frac{e}{4\pi \epsilon_0 \epsilon_n}\int_{0}^{\infty}J_{0}(k\rho)\left[A_2 (k)e^{kz}+B_2 (k)e^{-kz} + e^{-k|z-z_n|}\right]dk
\end{equation}
\end{widetext}
As for the lower (upper) h-BN regions, the solution is 
\begin{widetext}
\begin{equation}\label{solutionhBN}
\Phi_{n}(\rho,z) = \frac{e}{4\pi \epsilon_0 \epsilon_n}\int_{0}^{\infty}J_{0}(k\rho)\left[A_{1(3)} (k)e^{kz}+B_{1(3)} (k)e^{-kz}\right]dk.
\end{equation}
\end{widetext}

At the lower interface between two regions with different dielectric constants in Fig. \ref{fig:modelsystem}, namely, between h-BN and $N$-BP, the functions for each slab, given by Eqs.(\ref{solution}) and (\ref{solutionhBN}), must obey the following boundary conditions:
\begin{center}
\begin{equation}
\phi_{n,m}^{1}(\rho,z=d_1) = \phi_{n,m}^{2}(\rho,z=d_1), 
\end{equation}
\end{center}
\begin{center}
\begin{equation}
\epsilon_{1}\frac{\partial \phi_{n,m}^{1}(\rho,z=d_1)}{\partial z} = \epsilon_{2}\frac{\partial \phi_{n,m}^{2}(\rho,z=d_1)}{\partial z},
\end{equation}
\end{center}
where $d_1$ stands for the location of the interface between the two regions and the upper index identifies the solution at the regions 1 (lower h-BN layer) and 2 ($N$-BP), with dielectric constants $\epsilon_{1} = 4.5 \epsilon_0$ and $\epsilon_{2} = 10 \epsilon_0$, respectively. Similar boundary conditions between regions 2 and 3 (upper h-BN layer) are straightforwardly obtained.

\begin{figure}[!t]
\centerline{\includegraphics[width = 0.7\linewidth]{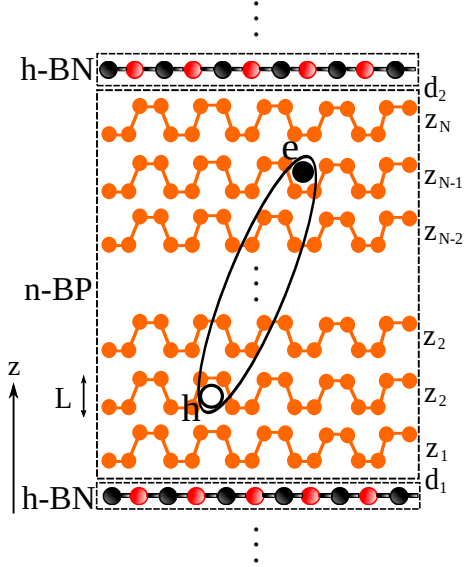}}
\caption{(color online) Sketch of the system under investigation, consisting of a stack of $N$ BP layers along the $z$-direction, encapsulated in h-BN. Electron (e) and hole (h) are localized in independent BP layers, assumed to have a thickness $L$.} \label{fig:modelsystem}
\end{figure}

The ETM method will be applied here in a three region problem, in which hexagonal boron nitride (h-BN) stands for regions 1 and 3, while few-layer black phosphorus ($N$-BP) stands for region 2, where electron and hole are assumed to lay. Region 2 embeds $N$ layers of BP, and we do not need to apply boundary conditions recursively between BP layers, as one would in a heterostructure, since these layers are composed by the same material. 

Assume an electron at layer $n$ generating a potential for a hole at layer $m$. In order to obtain the electron-hole interaction potential in the BP region by Eq. (\ref{solution}), one needs to find the amplitudes $A_2 (k)$ and $B_2 (k)$. These coefficients depend of the position of the electron in region 2, given by $z_n$. Therefore, those boundary conditions lead to a pair of equations, between regions 1 and 2, given by
\begin{equation}
A_{1}(k)e^{kd_1} = A_{2}(k)e^{kd_1}+B_{2}(k)e^{-kd_1} + e^{-k|d_1-z_n|},
\end{equation}
\begin{equation}
\epsilon_{1}A_{1}(k)e^{kd_1} = \epsilon_{2}A_{2}(k)e^{kd_1}-\epsilon_{2}B_{2}(k)e^{-kd_1} - s\epsilon_{2}e^{-k|d_1-z_n|},
\end{equation}
where $s = sgn(z-z_n)$ and $B_1(k)$ was made zero to avoid divergence in the $e^{-kz}$ term, since region 1 is assumed to extend towards $z \rightarrow -\infty$. Applying the same boundary conditions between regions 2 and 3 leads the following pair of equations:
\begin{equation}
A_{2}(k)e^{kd_2}+B_{2}(k)e^{-kd_2} + e^{-k|d_2-z_n|} = B_{3}(k)e^{-kd_2},
\end{equation}
\begin{equation}
\epsilon_{2}A_{2}(k)e^{kd_2}-\epsilon_{2}B_{2}(k)e^{-kd_2} -  s\epsilon_{2}e^{-k|d_2-z_n|} = -\epsilon_{3}B_{3}(k)e^{-kd_2},
\end{equation}
where $A_3(k)$ was made zero to avoid divergence in $e^{kz}$, since region 3 is assumed to extend towards $z\rightarrow +\infty$. This set of equations can be re-written in a matrix form as
\begin{widetext}
\begin{large}
\begin{equation}
\left(\begin{array}{cccc}
e^{kd_1}&-e^{-kd_1}&-e^{-kd_1} & 0 \\ 
\epsilon_{1}e^{kd_1} &-\epsilon_{2}e^{kd_1}& \epsilon_{2}e^{-kd_1} & 0\\
 0&e^{kd_2}&e^{-kd_2}&-e^{-kd_2} \\
 0&\epsilon_{2}e^{kd_2}&-\epsilon_{2}e^{-kd_2}&\epsilon_{3}e^{-kd_2} \end{array}\right)\left(\begin{array}{c} A_1 \\ A_2 \\ B_2 \\ B_3 \end{array}\right) =\left(\begin{array}{c}e^{-kd_1} \\-s\epsilon_{2}e^{-kd_1} \\ -e^{-kd_2} \\ s\epsilon_{2}e^{-kd_2}\end{array}\right). 
\end{equation}
\end{large}
\end{widetext}
By solving this matrix equation, one obtains $A_2$ and $B_2$, which are thus substituted back in Eq. (\ref{solution}) in order to obtain the electrostatic potential generated by the electron at $z_n$ for any position $z$ along the $N$-BP stack. Just like the electron, the hole also assume only discrete positions in $z$, which represent the middle of BP layers, as seen in Fig. \ref{fig:modelsystem}. Therefore, the actual electron-hole interaction potential $\phi_{n,m}(\rho)$ is obtained just by taking $\phi_{n,m}(\rho) = \Phi_n(\rho,z \rightarrow z_m) $ in Eq. (\ref{solution}).

Notice that the term in brackets in Eq. (\ref{solution}) can be re-written as
\begin{equation}
\epsilon_{m,n}^{eff}(k) = \frac{1}{A_2 (k)e^{kz_m}+B_2 (k)e^{-kz_m} + e^{-k|z_m-z_n|}},
\end{equation}
which is thus interpreted as an effective dielectric screening for this system. We verified that this function converges to the Rytova-Keldysh linear approximation in the limit of thin slabs and $n = m$, and the electron-hole interaction potential given by this procedure for $n \neq m$ has been demonstrated in Ref. [\onlinecite{Lucas2017}] to converge to the Coulomb interaction between charges in separate layers as the distance between electron and hole increases.

\section*{Appendix: Inter-layer hopping parameters}

As illustrated in Fig. 1(b) of the main manuscript, as the number of layers increase, the conduction (valence) band edge split into sub-bands, $N$ for $N$-BP. This is analogous to the splitting of energies observed e.g. in coupled quantum wells. In that case, for a double well, one can write a 2$\times$2 matrix with diagonal terms $E_0$ (e.g. the quantum well ground state energy) and off diagonal terms $t$, so that its eigenvalues are $E_0 \pm t$. The hopping parameter $t$ is then adjusted as to match the actual eigenvalues of the $2$-layer BP problem, obtained by diagonalization of the tight-binding Hamiltonian of multi-layer BP proposed in Ref. [\onlinecite{Duarte2017}], which is shown to yield accurate band structures as compared to ab initio calculations. \cite{Rudenko2015} One can then infer the result for three layers by writting a 3$\times$3 Hamiltonian matrix, with the same diagonals and off-diagonals, whose eigenvalues are $E_0, E_0 \pm \sqrt{2} t$. For an arbitrary number of layers $N$, such Hamiltonian matrix is tridiagonal and assumes the so called Toeplitz form. Eigenvalues of the tridiagonal Toeplitz matrix are given by
\begin{eqnarray}\label{1DTM}
E_{n} = E_0 + 2t \cos[n \pi /(N+1)],
\end{eqnarray} 
where the $n = 1,2,..,N$ is the energy state index. This approach is followed here for the description of conduction ($t = t_e$, $E_0 = E_{0e}$, where $E_{0e} = E_{e}-E_{h}$ is the gap of monolayer BP) and valence ($t = t_h$, $E_0 = E_{0h}$ = 0) band edges. Comparison with full TB calculations of multi-layer BP shows that this simple approach quantitatively match the TB results at the $\Gamma$-point of the first Brillouin zone if we take $t_{e(h)} = 0.299$ eV (0.499 eV) and $E_{0e}$ = 2.12 eV. \cite{Duarte2017} In fact, the effective gap of BP as a function of the number of layers as obtained by this method agrees well with experimentally observed ones.\cite{AndreyTony}

\section*{Appendix: Bulk limit of the sub-band exciton Hamiltonian}

It is instructive to verify how the sub-band exciton approach proposed in the main manuscript for $N$-BP behaves as $N \rightarrow \infty$ (bulk limit). In this case, in the absence of in-plane energy contributions, the set of eigenenergies for each quasi-particle, given by the eigenvalues of the Toeplitz matrix Hamiltonian [see Eq. (6) of the main manuscript], would form a band along the $k_z$ direction, $E_{e(h)}(k_z) = E_{0e(h)} + 2t_{e(h)} \cos\left(k_z d\right)$, within the Brillouin zone $-\pi/d < k_z < \pi/d$. In the vicinity of $k_z = 0$, this expression can be approximated by a Taylor series expansion $E_{e(h)}(k_z) \approx E_{0e(h)} + 2t_{e(h)} + t_{e(h)} d^2 k_z^2$, and this parabolic approximation for the quasi-particle energy along $k_z$ can be compared to an effective mass model to yield $m^z_{e(h)} = \hbar^2 \big/ 2t_{e(h)} d^2$ as the effective mass in $z$-direction for that quasi-particle $e(h)$. Comparison with results in the literature for effective masses in bulk BP agrees well for the values of $d = 0.52$ nm and $t_{e(h)}$ assumed here. \cite{Morita}

\section*{Appendix: Electron-hole separation along the BP layers}

\begin{figure}[!t]
\centerline{\includegraphics[width =\linewidth]{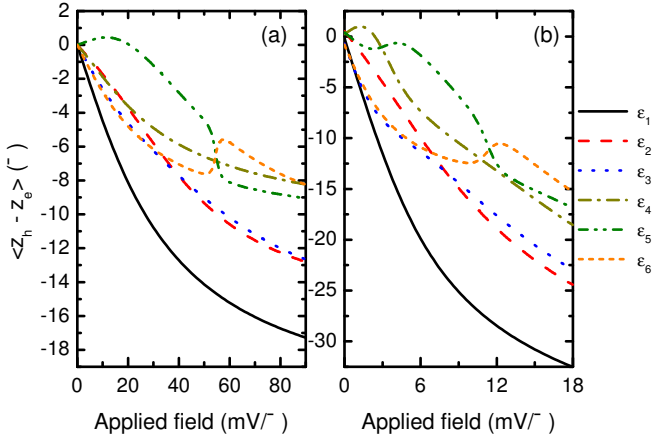}}
\caption{(color online) Electron-hole separation for the six lowest lying energy states as a function of the applied field in (a) 5-BN and (b) 10-BN.} \label{fig:dipoles}
\end{figure}

The electric field, applied perpendicularly to the BP layers, separates the electron and hole in oposite sides of the BP slab. Its important to investigate how efficient this field-induced electron-hole separation is in each exciton state as the field intensity increases. This information is important e.g. in the context of exciton-exciton interactions, where the dipole moment of the excitons play an important role. For the first six low-lying exciton states in (a) 5-BP and (b) 10-BP, Fig. \ref{fig:dipoles} shows the average electron-hole separation as a function of the applied field, calculated by 
\begin{equation}
\langle z_h - z_e \rangle = \sum_{i = 1}^N \sum_{j = 1}^N |\psi_{i,j}|^2 d_{i,j},
\end{equation}
where $\psi_{i,j}$ is the wave function component related to an electron in layer $i$ and a hole in layer $j$, as defined in the main manuscript, and $d_{i,j} = z_j - z_i$. In 5(10)-BP, the maximum electron-hole separation, given by the distance between the first and last BP layers in the $N$-BP stack, is 26 \AA\, (52 \AA\,). For fields as high as 90 mV/\AA\, (18 mV/\AA\,), the ground state reaches $\langle z_h - z_e \rangle$= 17 \AA\, (32.5  \AA\,). Strong fields are expected to change the band structure of few-layer BP, \cite{Kim2015} namely, by closing its gap, therefore, we restrict our calculations to lower fields as those in Fig. \ref{fig:dipoles}. Excited states exhibit crossings that result from the anti-crossings between their energies, as observed in Fig. 2 of the main manuscript. At high fields, the dipoles of the $\varepsilon_2$ and $\varepsilon_3$ states, predomintantly $E_{12}$ and $E_{21}$ sub-band states (see main manuscript), are similar and smaller (in modulus) as compared to one of the ground state (predominantly $E_{11}$).

\section*{Appendix: Comparison with photo-current experiments on few-layer transition metal dichalcogenides}

The excitonic lifetime has fundamental importance in opto-electronic applications of semiconductor materials. In photodetectors, for instance, a long radiative lifetime is desirable, since it is required that the electrons and holes created in the excitonic light absorption process are collected by the gates of the device before exciton recombination. The understanding and control of the dark and bright sub-band excitons provided by the model proposed here can give insights into the role of excitons in photoresponse experiments with multi-layer BP, as we will discuss in what follows.

In a recent experiment, \cite{Massicote2016} the use of a few-layer semiconductor TMD, namely WSe$_2$, as a photodetector has been proposed, where tuning of photoresponse by the number of layers, as well as by an external perpendicular electric field, was demonstrated. It would be interesting to compare the theoretical predictions for the similar case of few-layer BP studied here to these experimental results. The photoresponse in this system is described by the ratio $\Gamma = 1/\tau$, where $\tau^{-1} = \tau_r^{-1} + (\tau_d + \tau_s)^{-1}$ is a characteristic time constant, consisting of the radiative lifetime $\tau_r$, drift time $\tau_d$ and a bias-independent dissociation time $\tau_s$. The time constant $\tau$ was experimentally found to decrease with the external field and increase as $L^{1.9}$ with the width of the WSe$_2$ stack. The scaling exponent of $\approx 2$ in $L$ provides hints on the physical process: if excitons formed in the middle WSe$_2$ layer have their charge carriers moving towards the gates on the top and bottom of the WSe$_2$ with drift velocity $v_d$, the time for the photoresponse process is roughly $\tau = L/2v_d$, assuming that the drift process is the limiting timescale, compared to both radiative recombination and the typically much smaller dissociation time $\tau_s$. The drift velocity depends on the perpendicular electric field as $v_d = \mu F$, where $\mu$ is the mobility in direction perpendicular to the layers. For a fixed electrostatic bias across the layers $V_B$, the electric field depends on the thickness of the slab approximately as $F \approx V_B/L$, hence, the time constant due to a drift process is $\tau \approx \tau_d \approx L^2/2\mu V_B$. This simplistic model is demonstrated to describe well the experimental results for moderate values of the ratio $L^2/V_B$, but a higher bound for the time constant is observed at large $L$ (or, equivalently, lower $V_B$), which is interpreted as being due to the fact that the photocurrent efficiency is limited by the recombination time $\tau_r$.\cite{Massicote2016} In BP, the high electron and hole hopping parameters suggest a strong inter-layer coupling, which thus raises questions on how the layer dependence of $\tau_r$ and $\tau_d$ compare in few-layer BP.

\begin{figure}[!h]
\centerline{\includegraphics[width = 0.8\linewidth]{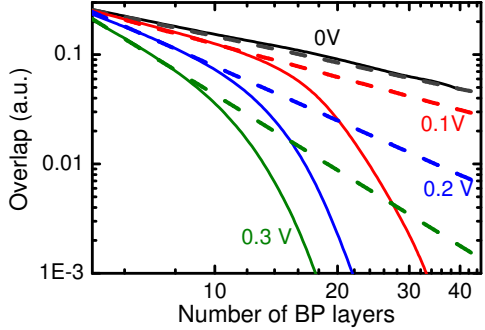}}
\caption{(color online) Electron-hole overlap as a function of the number of layers, for the ground state ($E_1$), at different values of the bias: 0 V (black), 0.1 V (red), 0.2 V (blue), and 0.3 V (green). Dashed lines are power law fittings of the low-$N$ results, $0.93N^{-0.8}$ (0 V), $1.25N^{-1.0}$ (0.1 V), $3.5N^{-1.65}$ (0.2 V), $8.6N^{-2.3}$ (0.3 V).} \label{fig:figbias}
\end{figure}

By exploring the link between the oscillator strength of excitonic states and the inverse exciton lifetime, we verify how the former depends on the BP thickness, as well as on an applied bias. Figure \ref{fig:figbias} shows the electron-hole overlap of excitonic transitions as a function of the number of BP layers under a bias with fixed value. The energies corresponding to these excitonic states are found in Fig. 2 of the main manuscript.

Assuming a BP-based photodetector operating at the frequency of the optical gap (i.e. with energy $\varepsilon_1 = E_{11}$, see Fig. 1(a) of the main manuscript), we now focus on the dependence of the ground state overlap on the number of layers. For a small number of layers, the contribution due to the drift process to the photoresponse time is small and thus dominates the charge carriers dynamics, being limited only by the transfer time $\tau_s$ of electrons and holes at the gates. In this case, the oscillator strength for small $N$ ($L \propto N$) is required to be small enough to produce an exciton lifetime $\tau_r$ larger than the drift time, so that radiative recombination does not limit the photoresponse. Figure \ref{fig:figbias} shows that, as the number of layers increases, the oscillator strength decreases, which leads to even \textit{longer} exciton lifetimes. In the absence of bias, for instance, the overlap decreases approximately as $0.93N^{-0.8}$ (gray dashed line), and it decreases at an even higher power of $N$ in the presence of bias, see Fig. \ref{fig:figbias}. For comparison, in WSe$_2$,\cite{Massicote2016} $\tau_r$ is shown to increase roughly linearly with the thickness, thus suggesting a $\approx N^{-1}$ dependence of the overlap on the number of layers. 

Notice that the drift time is expected to increase roughly as $N^{2}$, while the ground state exciton lifetime at low $V_B$ increases at a much lower rate, $N^{0.8}$. Consequently, the drift time can overcome the exciton lifetime at a critical thickness, where photocurrent is limited by recombination losses, but this will depend on how these two time scales compare for small $N$. In practice, assuming the exciton lifetime $\tau_r \approx \gamma N^{0.8}$ and the drift time $\tau_d = d^2N^2\big /2\mu V_B$, such critical number of layers would be given by $N_c \approx (2\mu V_B \gamma/d^2)^{0.83}$. A proper evaluation of $\gamma$ would require either experimental measurements or a more detailed calculation of the radiative recombination process, which is beyond the scope of this paper. Nevertheless, a rough estimate can be made as follows: the lifetime of monolayer BP has been measured to be $2.2 \times 10^{-10}$s, \cite{yang2015} therefore, one can assume this value to be $\approx \gamma$. For a small bias, e.g. $V_B = 10$ mV, and assuming that the mobility in the vertical direction across the BP layers is much lower than that in the in-plane directions, $\mu \approx 0.1$ cm$^2$/Vs, one obtains a critical number of BP layers $N_c \approx 69 $ ($\approx$ 35 nm thick BP), above which the radiation losses become relevant. Results from this same approach using the parameters for WSe$_2$, where out-of-plane mobility is much lower $\mu \approx 0.01$ cm$^2$/Vs, are in good agreement with the experimental observations in Ref. \onlinecite{Massicote2016}, where significant radiation losses are verified for a lower critical number of layers. Since the exponents in Fig. \ref{fig:figbias} are significantly higher for a bias of e.g. 0.3~V, a much higher value of $N_c$ is expected in actual biased samples and, consequently, radiation losses are not expected to play an important role even for thick BP slabs. Notice that our predictions on the fast carriers drift in BP are based on a very conservative account of the out-of-plane carrier mobility, since the actual values for this material have been calculated to be much higher, $\mu \approx 400-500$ cm$^2$/Vs, \cite{Morita} which will lead to even higher $N_c$.

Let us apply the model proposed here for the photoresponse rate of multi-layer WSe$_2$, in order to verify its agreement with the known experimental observations. \cite{Massicote2016} We seek for a theoretical prediction the value of $V_B/L^2$ where the radiative recombination process dominates over the drift process in photoresponse, considering three values of WSe$_2$ thickness: 2.2 nm, 7.4 nm and 28 nm, which correspond, respectively, to $N$ = 3, 11, and 44. (assuming the usual value for inter-layer distance $d$ = 0.65 nm) We take the out-of-plane mobility of WSe$_2$ as $\mu \approx 0.01$ cm$^2$/Vs for both electrons and holes. Also, we assume a $\approx \gamma N^{1.1}$ dependence of $\tau_r$ on the number of layers, which is inferred by the experimental values $\tau_r$ = 40 ps, 130 ps, and 750 ps, for $N$ = 3, 11, and 44, respectively. This also yields $\gamma = 10$ ps, in the same order of magnitude as in previous reports. \cite{Robert2016} Finally, we are interested in the value of $V_B/L^2$ for which $\tau_r < \tau_d$, i.e. $\gamma N^{1.1} <  {(dN)}^2/2 \mu V_B$, which can be re-written as $V_B/L^2 < 1/2 \mu \gamma N^{1.1}$. For the values of the constants provided here, we find that the recombination process will dominate over the drift one at $V_B/L^2 < $ 1.49$\times 10^{-2}$, 3.58 $\times 10^{-3}$, and 7.78 $\times 10^{-4}$ V nm$^{-2}$, for $L$ = 2.2 nm, 7.4 nm and 28 nm. These results lay in the same order of magnitude as the values below which the predictions from the diffusive transport model, which describes the drift process, are no longer accurate e.g. in Fig. 4a of Ref. [\onlinecite{Massicote2016}], thus supporting the validation of our analysis.

A comparison between the exciton lifetime and the drift time of charge carriers across the layers in the out-of-plane direction suggest that the latter process overcomes the radiative decays even for thick multi-layer BP slabs. This is in contrast to multi-layer WSe$_2$ systems, \cite{Massicote2016} where photoresponse efficiency of thick samples were demonstrated to be limited by radiative losses. The disparity between these two materials lies in the higher mobility of BP in the out-of-plane direction and stronger inter-layer coupling, as compared to the one in TMDC.


\begin{thebibliography}{apsrev}


\bibitem{Bridgman} P. W. Bridgman, J. Am. Chem. Soc. 36, 7, 1344 (1914).

\bibitem{Keyes} Robert W. Keyes, Phys. Rev. 92, 580 (1953).

\bibitem{Andres} Andres Castellanos-Gomez, J. Phys. Chem. Lett. 6, 4280 (2015).

\bibitem{Morita} A. Morita, Appl. Phys. A 39, 227 (1986).

\bibitem{Avouris} P. Avouris,T. Heinz and T. Low, \textit{2D Materials} (Cambridge University Press, 2004).

\bibitem{chaves2020bandgap} A Chaves, et al., NPJ 2D Materials and Applications 4, 29 (2020).

\bibitem{PBP} Tony Low, Rafael Rold\'an, Han Wang, Fengnian Xia, Phaedon Avouris, Luis Mart\'in Moreno, and Francisco Guinea, Phys. Rev. Lett. 113, 106802 (2014).

\bibitem{LiuHan} Han Liu, Adam T. Neal, Zhen Zhu, Zhe Luo, Xianfan Xu, David Tom\'anek and Peide D. Ye, ACS Nano 8, 4033 (2014).

\bibitem{BPfet} Likai Li, Yijun Yu, Guo Jun Ye, Qingqin Ge, Xuedong Ou, Hua Wu, Donglai Feng, Xian Hui Chen, and Yuanbo Zhang, Nature Nanotech. 9, 372 (2014).

\bibitem{CastellanosGomez} Andres Castellanos-Gomez, Leonardo Vicarelli, Elsa Prada, Joshua O Island, K. L. Narasimha-Acharya, Sofya I. Blanter, Dirk J. Groenendijk, Michele Buscema, Gary A. Steele, and J. V. Alvarez,  2D Mater. 1, 025001 (2014).

\bibitem{LowT} Tony Low, A. S. Rodin, A. Carvalho, Yongjin Jiang, Han Wang, Fengnian Xia, and A. H. Castro Neto, Phys. Rev. B 90, 075434 (2014).

\bibitem{RuPen} Ruoming Peng, Kaveh Khaliji, Nathan Youngblood, Roberto Grassi, Tony Low, and Mo Li, Nano Letters 17, 6315 (2017).

\bibitem{YoungBlood} N. Youngblood, C. Chen, S. Koester and Mo Li, Nature Photon. 9, 247 (2015).

\bibitem{WW} William S. Whitney, Michelle C. Sherrott, Deep Jariwala, Wei-Hsiang Lin, Hans A. Bechtel, George R. Rossman, and Harry A. Atwater, Nano Letters 17, 78  (2017).

\bibitem{JaiSingh} Jai Singh, \textit{Optical Properties of Condensed Matter and Applications} (Wiley, 2006).

\bibitem{SLatini} S. Latini, T. Olsen, and K. S. Thygesen, Phys. Rev. B 92, 245123 (2015).

\bibitem{Alexey2014} Alexey Chernikov, Timothy C. Berkelbach, Heather M. Hill, Albert Rigosi, Yilei Li, Ozgur Burak Aslan, David R. Reichman, Mark S. Hybertsen, and Tony F. Heinz, Phys. Rev. Lett. 113, 076802 (2014).


\bibitem{Vinod2019} Jie Gu, Biswanath Chakraborty, Mandeep Khatoniar, and Vinod M. Menon, Nature Nanotechnology 14, 1024 (2019).

\bibitem{GuoweiAndrey} Guowei Zhang, Andrey Chaves, Shenyang Huang, Fanjie Wang, Qiaoxia Xing, Tony Low and Hugen Yan, Science Advances 4, eaap9977 (2018).

\bibitem{AndreyReview} Tony Low, Andrey Chaves, Joshua D Caldwell, Anshuman Kumar, Nicholas X Fang, Phaedon Avouris, Tony F Heinz, Francisco Guinea, Luis Martin-Moreno, and Frank Koppens, Nature materials 16, 182 (2017).

\bibitem{Li2016} Likai Li, Jonghwan Kim, Chenhao Jin, Guo Jun Ye, Diana Y. Qiu, Felipe H. da Jornada, Zhiwen Shi, Long Chen, Zuocheng Zhang, Fangyuan Yang, Kenji Watanabe, Takashi Taniguchi, Wencai Ren, Steven G. Louie, Xian Hui Chen, Yuanbo Zhang, and Feng Wang, Nature Nanotechnology 12, 21 (2017).

\bibitem{AndreyTony} Guowei Zhang, Shenyang Huang, Andrey Chaves, Chaoyu Song, V. Ongun \"{O}z\c{c}elik, Tony Low and Hugen Yan, Nature Communications 8, 14071 (2017).

\bibitem{Partoens} L. L. Li, B. Partoens, and F. M. Peeters, Phys. Rev. B 97, 155424 (2018).

\bibitem{DArf} J. D. S. Forte, D. J. P. de Sousa, and J. M. Pereira Jr, Physica E: Low-dimensional Systems and Nanostructures 114, 113578 (2019).

\bibitem{Rytova} N. S. Rytova, Proc. MSU, Phys., Astron. 3, 30 (1967).

\bibitem{Keldysh} L. V. Keldysh, \textit{JETP Lett.} 29, 658 (1979).

\bibitem{Lucas2017} L. S. R. Cavalcante, A. Chaves, B. Van Duppen, F. M. Peeters, and D. R. Reichman, Phys. Rev. B 97, 125427 (2018).

\bibitem{Island2015} Joshua O. Island, Gary A. Steele, Herre S. J. van der Zant, and Andres Castellanos-Gomez, 2D Materials 2, 011002 (2015).

\bibitem{Artel2017} Vlada Artel, Qiushi Guo, Hagai Cohen, Raymond Gasper, Ashwin Ramasubramaniam, Fengnian Xia, and Doron Naveh, npj 2D Materials and Applications 1, 6 (2017).

\bibitem{Duarte2017} D. J. P. de Sousa, L. V. de Castro, D. R. da Costa, J. M. Pereira, Jr., and Tony Low, Phys. Rev. B 96, 155427 (2017).


\bibitem{AndreyRC} A. Chaves, J. G. Azadani, V. O. \"{O}z\c{c}elik, R. Grassi, T. Low, Phys. Rev. B 98, 121302(R) (2018).

\bibitem{QW1} S. Glutsch, \textit{Excitons in low-dimensional semiconductors} (Berlin: Springer, 2004).

\bibitem{QW2} P. A. Belov, Journal of Physics: Conf. Series 1199, 012018 (2019).


\bibitem{Elliot} R. J. Elliot, Phys. Rev. 108, 1384 (1957).

\bibitem{Rudenko2015} A. N. Rudenko and M. I. Katsnelson, Phys. Rev. B 89, 201408(R) (2014).

\bibitem{Kim2015} J. Kim, S. S. Baik, S. H. Ryu, Y. Sohn, S. Park, B.-G. Park, J. Denlinger, Y. Yi, H. J. Choi, and K. S. Kim, Science \textbf{349}, 723 (2015).

\bibitem{molas2017} Maciej R. Molas, Karol Nogajewski, Artur O. Slobodeniuk, Johannes Binder, Miroslav Bartosa and Marek Potemski, Nanoscale 9, 13128 (2017).

\bibitem{zhao2013} Weijie Zhao, Zohreh Ghorannevis, Leiqiang Chu, Minglin Toh, Christian Kloc, Ping-Heng Tan, and Goki Eda, ACS Nano 7, 791 (2013).

\bibitem{mak2010} Kin Fai Mak, Changgu Lee, James Hone, Jie Shan, and Tony F. Heinz, Phys. Rev. Lett. 105, 136805 (2010).


\bibitem{Zipfel2018} Jonas Zipfel, Johannes Holler, Anatolie A. Mitioglu, Mariana V. Ballottin, Philipp Nagler, Andreas V. Stier, Takashi Taniguchi, Kenji Watanabe, Scott A. Crooker, Peter C. M. Christianen, Tobias Korn, and Alexey Chernikov, Phys. Rev. B 98, 075438 (2018).

\bibitem{Stier2016} Andreas V. Stier, Kathleen M. McCreary, Berend T. Jonker, Junichiro Kono, and Scott A. Crooker, Nat. Commun. 7, 10643 (2016). 

\bibitem{Stier2018} A. V. Stier, N. P. Wilson, K. A. Velizhanin, J. Kono, X. Xu, and S. A. Crooker, Phys. Rev. Lett. 120, 057405 (2018).

\bibitem{Goryca2019} M. Goryca, J. Li, A. V. Stier, T. Taniguchi, K. Watanabe, E. Courtade, S. Shree, C. Robert, B. Urbaszek, X. Marie, and S. A. Crooker, Nature Communications 10, 4172 (2019).

\bibitem{Chen2019Luminescent} Shao-Yu Chen, Zhengguang Lu, Thomas Goldstein, Jiayue Tong, Andrey Chaves, Jens Kunstmann, L. S. R. Cavalcante, Tomasz Wo\'zniak, Gotthard Seifert, D. R. Reichman, Takashi Taniguchi, Kenji Watanabe, Dmitry Smirnov, and Jun Yan, Nano Lett. 19, 2464 (2019).

\bibitem{yang2015} Jiong Yang, Renjing Xu, Jiajie Pei, Ye Win Myint, Fan Wang, Zhu Wang, Shuang Zhang, Zongfu Yu, and Yuerui Lu, Light: Science and Applications 4, e312 (2015).

\bibitem{Robert2016} C. Robert, D. Lagarde, F. Cadiz, G. Wang, B. Lassagne, T. Amand, A. Balocchi, P. Renucci, S. Tongay, B. Urbaszek, and X. Marie, Phys. Rev. B 93, 205423 (2016).

\bibitem{Massicote2016} M. Massicotte, P.Schmidt, F. Vialla, K. G. Sch\"adler, A. Reserbat-Plantey, K. Watanabe, T. Taniguchi, K. J. Tielrooij, and F. H. L. Koppens, Nat. Nanotech. 11, 42 (2016).


\end{thebibliography}
\end{document}